\documentclass{rmaa}

\title{A revised calibration of the $M_V-W(OI~7774)$ relationship using Hipparcos data: Its application to Cepheids and evolved stars\altaffilmark{1}}

\author{A. Arellano Ferro\altaffilmark{2}, Sunetra Giridhar\altaffilmark{3}
  and E. Rojo Arellano\altaffilmark{2}}

\altaffiltext{1}{Based on observations collected at the San Pedro M\'artir Observatory, Mexico}
\altaffiltext{2}{Instituto de Astronom\'{\i}a, Universidad Nacional Aut\'onoma de M\'exico.\\
 Apdo Postal 70-264, M\'exico D. F. CP 04510, Mexico}
\altaffiltext{3}{Indian Institute of Astrophysics, Bangalore 500034, India}

\fulladdresses{
\item A. Arellano Ferro and E. Rojo Arellano: Instituto de Astronom\'{\i}a,
  Universidad Nacional Aut\'onoma de M\'exico, Apdo Postal 70-264, M\'exico D. F. CP 04510, Mexico(armando@astroscu.unam.mx; cefeida@yahoo.com)
\item Sunetra Giridhar: Indian Institute of Astrophysics, Bangalore 500034, India.
  (giridhar@iiap.ernet.in)
  }

\shortauthor{A. Arellano Ferro et al.}
\shorttitle{The $M_V-W(OI~7774)$ relationship revised}

\SetVolume{XX} \SetFirstPage{1} \SetYear{2002}
\ReceivedDate{July 2002}
\AcceptedDate{------} 

\resumen{Se ha calculado una nueva calibraci\'on de la correlaci\'on $M_V-W(OI~7774)$
usando mejores estimaciones de enrojecimientos y
distancias de 27 estrellas ca-libradoras de tipos A-G, basadas en los paralajes y 
movimientos propios de los cat\'alogos Hiparcos y Tycho. La nueva calibraci\'on 
predice magnitudes absolutas con precisiones de $\pm$ 0.38 mag. para
una muestra que cubre el rango de $-9.5$ a $+0.35$ mag. El t\'ermino de color, usado en
calibraciones anteriores, no fue encontrado significativo y fue eliminado.
Las variaciones del triplete OI7774  a lo largo del ciclo de pulsaci\'on en la 
cefeida cl\'asica SS Sct fueron estudiadas. Calculamos  
una correcci\'on dependiente de la fase, a 
la intensidad del triplete de OI obtenida en fase al azar en Cefeidas, para 
determinar su magnitud absoluta media a traves de la relaci\'on $M_V-W(OI~7774)$. Despu\'es de aplicar
esta correcci\'on a una muestra de 31 Cefeidas, pudimos incrementar la
lista de calibradoras a 58. La desviaci\'on estandar de la calibraci\'on usando 
la muestra combinada es comparable a la obtenida a partir de las 27 calibradoras 
primarias,
lo que indica que el m\'etodo permite obtener luminosidades medias de Cefeidas a partir
de observaciones de OI7774 obtenidas al azar. Las calibraciones
fueron aplicadas a un grupo de estrellas evolucionadas para posicionarlas en el diagrama
H-R.}

\abstract{A new calibration of the $M_V-W(OI~7774)$ relationship has 
been calculated  using better estimates of reddenings and distances to a 
 sample of 27 calibrator stars of A-G spectral types, based on accurate parallaxes 
and proper motions from the Hipparcos and Tycho catalogues. 
The  present  calibration predicts absolute magnitude with accuracies of 
$\pm$ 0.38 mag.
for a sample covering a large range of $M_V$, from $-9.5$ to $+0.35$ mag.
 The colour term  included in a previous paper has been dropped since its
 inclusion was not bringing any significant improvement to the calibration.
The variation of the OI7774 feature in
 the classical cepheid SS Sct has been studied. We calculated a phase-dependent correction to random phase OI feature strengths 
in Cepheids, such that it  predicts mean absolute 
 magnitudes using the above calibration. After 
applying such correction, we could increase the list of calibrators to 58 by 
 adding $M_V$ and OI triplet strength data for 31 classical Cepheids.
  The standard error of the calibration using the composite sample was 
comparable to that obtained
from the primary 27 calibrators, showing that it is possible to calculate mean
Cepheid luminosities from random phase observations of the OI7774 feature. 
We use our derived calibrations to estimate $M_V$ for a set of
evolved objects to be able to locate their positions in
the H-R diagram.}
      
\keywords{Absolute Magnitudes, Trigonometrical Parallaxes
, OI~7774 line strengths}

\begin{document}

\maketitle

\section{Introduction}

 Positioning a star or a family of stars in the H-R diagram is fundamental
 to understand the structure and evolution of stars since it enables
 proper comparison with evolutionary tracks and computed isochrones.
 When supplemented by chemical composition data, one can get a much 
 deeper insight into the evolutionary processes that the star might
 have undergone before arriving at the present stage.
Though the temperature of the star can be estimated by different
 methods like photometry, scanner observations, shapes of hydrogen
 and helium lines, from excitation equilibrium of species like
 Fe, Cr, Ti etc. with increasing accuracy, estimating absolute 
 magnitudes is not always easy. For hot stars, the profiles of 
 helium and hydrogen are employed whereas for cool stars
 Mg II lines at 2796.3 $\AA$,  Ca II H and
 K lines at 3933 and 3967$\AA$ and Mg I triplet in 5167-5184$\AA$ 
 region are found to be good indicators of luminosity.
 A summary of spectral features that are good indicators of spectral types
 and luminosity types can be found in Jaschek \& Jaschek (1987).
 
For stars of spectral type A-F the OI triplet at $\lambda \lambda$
7771.954, 7774.177 and 7775.395 $\AA$ is found to be
 a very good indicator of luminosity.

The sensitivity of the OI7774 triplet to the stellar luminosity has been well
known since Merril (1925, 1934) noted striking strength differences
 of the feature among supergiants
and main sequence stars. Keenan \& Hynek (1950)
studied the variation of the feature with spectral type and proposed the use
of this feature as a luminosity indicator, noticing its
 large strength in  A and
F type stars. Osmer (1972) performed the first calibration of the OI7774 
triplet
in terms of absolute magnitudes using a photolectric approach
 for 10 F-type supergiants. After his  pioneering work, several 
 calibrations of the feature were carried out  spectroscopically and
photometrically (e.g. Baker 1974, Sorvari 1974, Kameswara Rao \& Mallik 1978,
 Arellano Ferro et al. 1989;
1991, Arellano Ferro \& Mendoza 1993  , Mendoza \& Arellano Ferro 1993,
Slowik \& Peterson 1993; 1995). A good compendium on the 
feature intensity across the H-R diagram can be found in the work 
of Faraggiana et al. (1988). In the paper by Arellano Ferro, Giridhar \& Goswami (1991)
  (paper I) a calibration of the
feature was  made using high resolution data and a discussion of the role of the
resolution is given. It was
  demonstrated  that the equivalent widths at low resolution can be
overestimated by as much as 30\% and that a better calibration of the $M_V-W(OI~7774)$
relationship can be obtained at the
 resolution $R\sim 18,000.$ for the large $M_V$  range ($-10$ to $+2$ mag).
We felt that a calibration covering a larger range in spectral type 
would have wider application. Though we have reduced calibration errors 
by measuring the $W(OI7774)$ feature at high resolution data, the 
$M_V$ data on the calibrators (mostly members
of clusters and associations) was not meeting the required  accuracy.

 Fortunately, great improvement has been made in the last decade in the
 determination of  distances and reddenings to the parent groups
 that contain the calibrator stars used in paper I. With accurate parallaxes
 and proper motions from Hipparcos (ESA 1997), the data on open clusters,
 such as number of confirmed members, mean proper motions and  parallaxes has
 vastly improved (e.g. Baumgardt, Dettbarn, \& Wielen (2000); 
Tadross 2001).  We, therefore, decided to redetermine $M_V$ for
 the calibrators and calculate a new $M_V-W(OI~7774)$ relationship
 that would help in determining $M_V$ for field stars more accuarately.
 We describe in section 2 our observational material.
  In sections 3 and 4 the list of calibrators
and their new $M_V$-values  are presented and the new calibration
 is discussed. In section 5 we discuss the behaviour of the OI~7774  feature 
 in classical Cepheids and explore the possibility of
 using them as  additional calibrators.
In section 6,  we calculate the
  luminosities for a group of selected evolved stars and their
position in the H-R diagram is given  using our estimated luminosities.
We summarize our findings in section 7.   

\section{Observations}
\label{sec:Observations}
 Most observations were carried out in August 2001 and January 2002, with
the 2.1m telescope of San Pedro Martir Observatory (SPM), Mexico, 
equipped with a
Cassegran Echelle spectrograph and a CCD Site SI003 of 1024$\times$1024 pixels.
 This instrument gives    a resolution of $\sim$18,000.
at 7774 $\AA$. Along with the stellar observations,
bias frames and He-Ar lamp spectra were obtainted to carry out
 background subtraction and wavelength calibration. All reductions were
 made using standard procedures and tasks contained in the IRAF package.

The observations of the cepheid $\zeta$ Gem were obtained with the 1.0m telescope of
 Vainu Bappu Observatory at Kavalur, India. This telescope is 
 equipped with a Coud\'e Echelle spectrograph giving a resolution of $18,000$.
 The spectra were recorded on a 
Thompson-CSF7H7882 CCD of 384$\times$576 pixels.                 

\section{The calibrators stars}
\label{sec:Calibrators}

We have  chosen from the calibrator stars employed in paper I (Table 1),
 only those objects  that are
observable from the latitude of SPM. We have retained in our present list of
calibrators only those stars for which we have a
new determination of W(OI7774), and/or a new $M_V$ estimated as discussed below.
Several calibrators are members of open
clusters or OB associations as listed by Arellano Ferro \& Parrao (1990)
(their Table 1). However, their absolute magnitudes have been recalculated as 
new proper motions studies can be used to confirm
their membership (Baumgardt, Dettbarn \& Wielen 2000) and new distances and
reddenings (Tandross 2001) are available for clusters. Accurate parallaxes
  and proper motions  are available even for some   field stars
(Hipparcos, ESA 1997) now. Therefore, the list of calibrators also contains
some field stars for which new values of $M_V$ have been estimated.

\subsection{The calibrators in clusters and associations}

 The accurate parallaxes and proper motions given in the Hipparcos catalogue,
 when combined with their ground based counterparts and radial velocities of 
known cluster and associations members, can lead to
 much improved values of the mean proper motions and parallaxes for a large
number of galactic clusters and associations. Baumgardt, Deltbarn \&
Wielen (2000) have derived these quantities for 205 open clusters.
These authors also give a list of confirmed and possible members of these
clusters. Their work also indicates downward revision in the distances
by about 12\% compared to the photometric estimates. This paper 
enabled us to further confirm the membership in clusters of the stars
used in the present work and to re-evaluate $M_V$.

\begin{table}[t]
\footnotesize{
\caption{Calibrator stars of the $M_V-W(OI~7774)$ relationship}                                                                 
\begin{center}
\begin{tabular}{rlllll}
\noalign{\smallskip} 
\hline       
\noalign{\smallskip}                                                              
\noalign{\smallskip}
\multicolumn{1}{l}{HD}&
\multicolumn{1}{l}{Sp.T.}& 
\multicolumn{1}{l}{W71}& 
\multicolumn{1}{l}{W74}&
\multicolumn{1}{l}{$M_V$}&
\multicolumn{1}{l}{$(b-y)_o$}\\
\noalign{\smallskip}
\hline  
\noalign{\smallskip}  

7927   & F0Ia & 0.855 & 2.221 & --8.76 & 0.111\\
9973   & F5Iab&  0.541 & 1.399 & --7.36 & 0.225\\
10494  & F5Ia & 0.612 & 1.578 & --7.34 & 0.215\\
14433  & A1Ia & 0.661 & 1.641 & --7.08 & 0.013\\
14535  & A2Ia & 0.496 & 1.198 & --5.97 & 0.124\\
17971  & F5Ia & 0.538 & 1.404 & --6.58 & 0.247\\
18391  & G0Ia & 0.611 & 1.438 & --6.6 & 0.949\\
20123  & G6Ib & 0.124 & 0.279 & --2.0 & 0.645\\
20902  & F5Ib & 0.374 & 1.020 & --4.9 & 0.274\\
31964  & F0Ia & 0.969 & 2.316 & --8.7 & 0.035\\
36673  & F0Ib & 0.449 & 1.249 & --5.1 & 0.116\\
48329  & G8Ib & 0.087 & 0.182 & --1.0 & 0.816\\
54605  & F8Ia & 0.691 & 1.750 & --7.97 & 0.355\\
62058  & G0Ia & 0.579 & 1.485 & --7.32 & 0.981\\
62345  & G8IIIa &0.042 & 0.084 & +0.35 & 0.541\\
65228  & F7II & 0.216 & 0.547 & --1.9 & 0.46\\
74180  & F0Ia &   & 2.273* & --9.0 & 0.094\\
75276 & F2Iab &   & 1.114* & --6.45 & 0.011\\
84441  & G1II & 0.117 & 0.289 & --1.31& 0.36\\
87283 & F0II &   & 1.017* & --4.01 & 0.113\\
90772 & F0Ia &   & 2.051* & --8.3 & 0.054\\
101947 & G0Ia &   & 1.757* & --7.9 & 0.439\\
102070 & G8IIIa & 0.040 & 0.1118 & --0.5 & \\
164136 & F2II & 0.286 & 0.753 & --2.73 & 0.253\\
194093 & F8Ib & 0.490 & 1.288 & --6.18 & 0.397\\
204867 & G0Ib  & 0.237 & 0.622 & --3.37 & 0.40\\
217476 & G0Ia & 0.910 & 2.174 & --9.2 & 0.674\\
            \noalign{\smallskip}  
            \hline  
            \noalign{\smallskip}  
\end{tabular}
\end{center}
* -- W74 values taken from paper I.
}
\end{table} 

{\it HD 7927 ($\phi$ Cas)}. It is a member of open cluster NGC 457.
Tadross (2001) estimates $E(B-V)$=0.5 and a distance of
2851 pc for the cluster,  leading to  $M_V$=--8.76.

{\it  HD 9973} belongs to the association Cas OB1  with distance
modulus of 12.4,  with adopted $E(B-V)$=0.54  Oestreicher \&
Schmidt-Kaler (1999) find  $M_V$=--7.36.

{\it HD 10494} is a member of open cluster NGC 654. Tadross (2001)
estimates $E(B-V)$=0.90 and a distance of 2483 pc for
NGC 654, thus $M_V$=--7.34.

{\it HD 14433} is a member of open cluster NGC 884 ($\chi$ Persei) for which
Tadross (2001) estimates $E(B-V)$=0.50 and a distance of 2483 pc.
Therefore $M_V$=--7.08.

{\it HD 14535} is also a member of NGC 884 ($\chi$ Persei). As for HD 14433 we estimated $M_V$=--5.97.

{\it HD 17971} belongs to IC 1848  with distance modulus of 11.81, with
adopted $E(B-V)$=0.76 Oestreicher \& Schmidt-Kaler (1999) find
$M_V$=--6.58.

{\it HD 18391} is listed as a possible member of h-$\chi$ Per by Schmidt (1984). From the reddenings and $M_V$ compilation of Arellano Ferro \& Parrao (1990) we adopt $M_V= -6.6$.

{\it HD 20902}($\alpha$ Per) is a member of the $\alpha$ Persei cluster. Using the distance and reddening to the cluster, Humpreys (1978) found $M_V$= --4.7. Using Hipparcos parallaxes Jaschek \& G\'omez 1998 have
  calculated  $M_V$ for some MK standards. For HD 20902 they give $M_V$=--4.9.

{\it HD 31964} belongs the associations Aur OB1 (Stothers 1972). From the reddenings and $M_V$ compilation of Arellano Ferro \& Parrao (1990) we adopt $M_V= -8.7$.

{\it HD 54605} belongs to Collinder 121 with distance modulus of 9.4,
with adopted
 $E(B-V)$=0.12 Oestereicher \& Schmidt-Kaler (1999) find $M_V$=--7.97.

{\it HD 62058} is  a member of open cluster NGC 2439. Tadross (2001) estimates $E(B-V)$=0.37 and a distance of
3669 pc for the cluster, leading to $M_V$=--7.32.

{\it HD 74180} is a member of Vel OB1 association (Humpreys 1978). Two recent estimates of
 the distance are available 1600 pc (Dambis, Melnik \& Rostorguev 2001)
 and 1750 (Cameron Reed 2000). The later author has
also provided the value of the total to selective absortion ratio $R=3.7$ for
the association and $E(B-V)$=0.478 for HD 74180. This leads to $M_V=-9.1$
and $M_V=-8.9$ respectively. We adopted $M_V=-9.0$

{\it HD 75276} is also a member of Vel OB1 (Humpreys 1978). Following the steps
taken for HD 74180 and adopting
 $E(B-V)$=0.315 (Cameron Reed 2000), it is found to give
$M_V=-6.6$ and $M_V=-6.3$ respectively for the two
distances. We adopted $M_V=-6.45$

{\it HD 87283} is  a member of open cluster NGC 3114 (Schmidt 1984) for which
Tadross (2001) estimated $E(B-V)$=0.50 and distance of 2483 pc.
  Therefore we adopted $M_V=-4.01$.

{\it HD 90772} is  a member of open cluster IC2581 (Lloyd Evans 1969).
 From the reddenings and $M_V$ compilation of Arellano Ferro \& Parrao (1990)
 we adopt $M_V= -8.3$.

{\it HD 101947} belongs to Stock 14, (Schmidt 1984).
From the reddenings and $M_V$ compilation of Arellano Ferro \& Parrao (1990)
 we adopt $M_V =-7.9$.

\subsection{The field calibrators}

We have included in the list of calibrators a few field stars whose
$M_V$ values  can be correctly estimated using Hipparcos data.
Hipparcos parallaxes alone can be used to derive distances
if the value of the parallax is larger than 5 times the error of the
parallax value.

{\it HD 20123}.  Using Hipparcos parallax Wallerstein, Machado-Pelaez,
\& Gonzalez (1999), have derived  $M_V$=--2.0

{\it HD 36673} is a field star with $E(B-V)$ =0.02 Oestreicher \&
Schmidt-Kaler (1999) find $M_V= -5.1$.

{\it HD 48329}. As in paper I we adopted $M_V=-1.0$ from parallax value.

{\it HD 62345}.  Using Hipparcos parallax Allende Prieto \& Lambert (1999),
have derived $M_V= +0.35 \pm 0.16$.

{\it HD 65228} has a paralax of 6.49 leading to a distance of 154 pc.
  With $E(B-V)$=0.057 given by Bersier (1996) one gets $M_V= -1.9$.

{\it HD 84441} has a large parallax of 13.01, indicating a distance of
76 pc. With this distace and $E(B-V)=-0.04$ from Arellano Ferro \& Parrao (1990)
 one gets $M_V= -1.31$.

{\it HD 102070} has parallax of 9.31 leading to a distance of 107 pc.
  With $E(b-y)=0.02$ or $E(B-V)=0.025$ given in Paper I,
  we find $M_V=-0.5$.

{\it HD 164136} has a parallax of 4.10 leading to a distance of 467 pc.
  With $E(B-V)=0.07$ (Bersier 1996) then $M_V= -2.73$.

{\it HD 194093} has a paralax of 2.14 leading to a distance of 244 pc.
  With $E(B-V)$=0.026 (Bersier 1996) leads to  $M_V= -6.18$.

{\it HD 204867} has a parallax of 5.33 leading to a distance of 187.6 pc. With $E(B-V)$=0.026 (Bersier 1996) one finds $M_V=-3.37$.

{\it HD 217476} is considered an hypergiant. Its log$(L_*/L_{\odot})$ is
estimated to be 5.6 by de Jager (1998), that leads to $M_V=-9.2$.

In Table 1 we sumarize the
newly calculated values of $M_V$ for each calibrator. We  also tabulate in
 Table 1, the individual strength of OI at 7771.95$\AA$ along with 
the strength of 7774 blend comprising the three 
components.  It should be noted that between the OI line at
7771.95$\AA$ and the next one at 7774.17 $\AA$, features of Fe I at 7772.59 $\AA$
and CN at 7772.9 $\AA$ are present. These lines are weak or non-existent
 in A type stars but become prominent in stars of spectral type G or later.
 The contributions from these lines could cause overestimation of
 triplet strengths in relatively cooler stars. However, for OI 7771.95 $\AA$
 the blue side of its profile remains largely unaffected, hence more
 accuarate estimate of its strength can be made.
 At our resolution, OI 7771.954$\AA$ was distinctly separated from the
 rest of the blend. In an attempt to get a calibration with smaller dispersion
 and possible extension of the calibration to the cooler stars we calculated
 two separate calibrations, one using only the first component at OI 7771.957
$\AA$ hereinafter called W71 and the other, using the combined strength 
of three components hereinafter called W74.
 The W74 value for those stars with a 
new estimate of $M_V$ but not observed in SPM has been adopted from paper I,
these cases are marked with an asterisk in Table 1 and plotted as circles
in Figs. 1b and 6b.
 It should be noted that the spectra used in paper I were also
 of resolution very similar to that of SPM spectra.
The $(b-y)_o$ values were calculated from Str\"omgren data and
 colour excesses given
in the literature (e.g. Arellano Ferro \& Parrao 1990 and paper I).

\begin{figure} [t]
\includegraphics[width=7.cm]{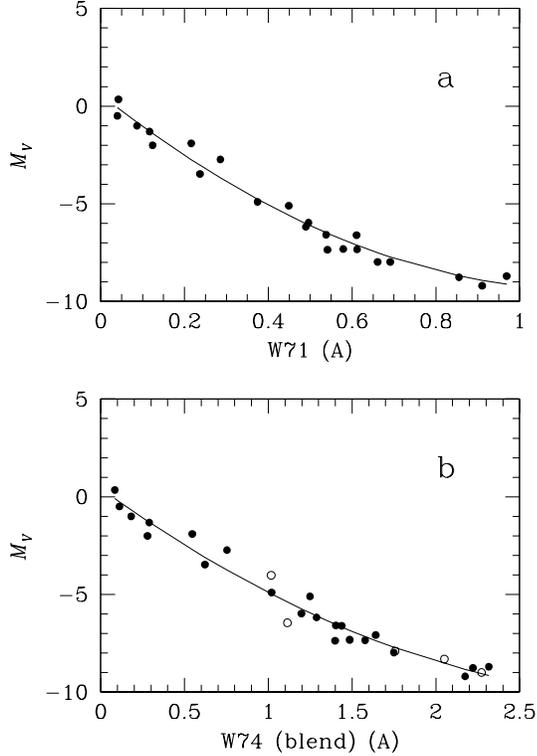}
\caption{Calibration of the equivalent widths of OI7771 (W71) and OI7774 (W74)
 (blend of three components) in terms of the absolute magnitude $M_V$. The solid lines are represented
by equations 1 and 2, which hold for the  absolute magnitude range --9.5 to +0.35 mag.
Open circles are the stars with adopted W74 from paper I.}
\end{figure}

\section{The $M_V-W(OI~7774)$ calibration}
\label{sec:Calibration}

Fig. 1 shows the distribution of the calibrators
 in equivalent width versus   absolute magnitude diagram 
for W71 and W74. The solid lines are the least square fits to
 the points and can be represented by the equations:

\begin{equation}
\begin{array}{@{}r@{}r@{~}r@{~}r@{~}r@{~}r@{}}
M_V = &\phantom{\pm}0.604& - 17.079&\rm{W71}& + 7.227&\rm{W71}^2, \\
 &\pm0.282&\phantom{0}\pm1.346& &\pm 1.383& 
\end{array}
\end{equation}

\begin{equation}
\begin{array}{@{}r@{}r@{~}r@{~}r@{~}r@{~}r@{}}
M_V = &\phantom{\pm}0.427&-6.234&\rm{W74}&+0.907&\rm{W74}^2. \\
 &\pm0.292&\phantom{0}\pm0.572& &\pm 0.243&
\end{array}
\end{equation}

The standard errors of the fits are 0.38 mag. for both calibrations. Figure
1b and equation 2  can be compared with Figure 3 and equation 1 of paper I, 
\begin{equation}
\begin{array}{@{}r@{}r@{~}r@{~}r@{~}r@{~}r@{}}
M_V =&\phantom{\pm}0.49&-6.33&\rm{W74}&+0.85&\rm{W74}^2, \\
 &\pm0.75&\phantom{0}\pm1.57& &\pm 0.68&
\end{array}
\end{equation}

\noindent
where the standard error of the fit was 1.5 mag.

\begin{figure} [t] 
\includegraphics[width=7.cm]{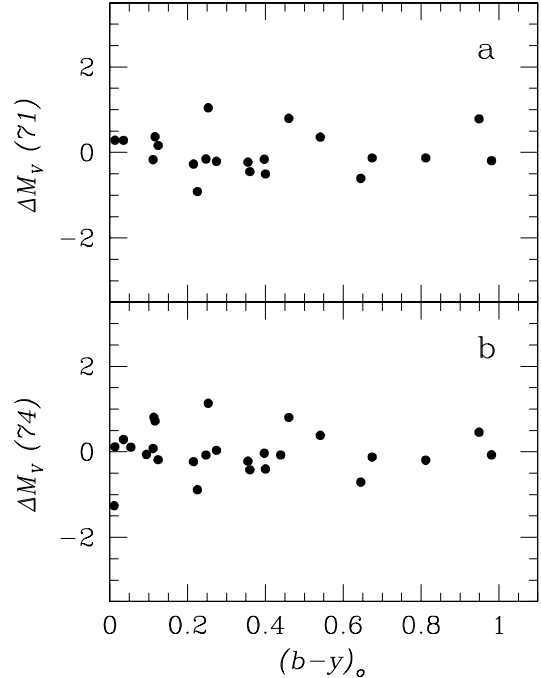}  
\caption{Residuals in Figure 1a and 1b in terms of the colour $(b-y)_o$. As no trend is seen, it is
evident that there is no significance of the colour in the $M_V-W(OI~7774)$ relationship.}  
\end{figure}

It is evident from the standard deviations of the fit and the reduced standard errors of        
the coefficients, that the new calibratios of equations 1 and 2 are better, by more
than a factor of three, than the old calibration, this is undoubtedly due to the 
improved values of $M_V$ calculated from the Hipparcos data.
As in paper I, we have  investigated the  effect of the
 temperature variation over the spectral types, represented by the colour 
$(b-y)_o$, in the relationship. The equations including the colour term are of the form
\begin{equation}
M_V =\rm{A}-\rm{B}\rm{W}+C\rm{W}^2 -D\it{(b-y)}_o 
\end{equation}


\noindent
with A$=1.015\pm0.403$, B=$-17.961\pm1.489$, C=$7.924\pm1.468$,
D$=-0.424\pm0.407$ for W71 and
A$=0.947\pm0.402$, B=$-6.544\pm0.603$, C=$0.986\pm0.246$,
D$=-0.734\pm0.414$ for W74. 
The standard deviations of the fit are 0.35 and 0.33 mag respectively, and,
although a bit smaller than for eqs. 1 and 2, the colour term does not appear
 to be  significant.  Further, the plots of the residuals 
  versus $(b-y)_o$ show  no trend with colour, as demonstrated in
 Fig. 2, thus the colour term was dropped from the calibration.
 Equations 1 and 2 hold good for absolute magnitudes in 
the large range --9.5 to +0.35 mag. and spectral types between A1 and G8.

In Fig 7-a examples of the OI7774 triplet are given for three
clibrator stars. They illustrate the considerable range of variation
of the OI7774 strength between low and  high luminosity stars.

\section{The Behaviour of OI(7774) feature in classical Cepheids}
\label{sec:Cepheids}

The  $M_V-W(OI~7774)$ calibrations discussed so far had been
 obtained  using  A-G giant and  supergiant calibrators 
with a large $M_V$ range of nearly 10 magnitudes. Therefore, they are 
expected to be valid for classical Cepheids too that are generally
F-G supergaints. 
But, it should be  noted that calibrating $M_V-W(OI~7774)$ independently
from Cepheids alone may not be possible due to their small range
 in $M_V$ (--1 to --5 mag.) and  variations in W74 as a function of phase.
 On the other hand, Cepheid   luminosities are believed
to be well determined from the Period-Luminosity (P-L) relation, hence
 they can be used to enlarge the list of calibrators. 

\begin{figure}[t]
\includegraphics[width=7.cm]{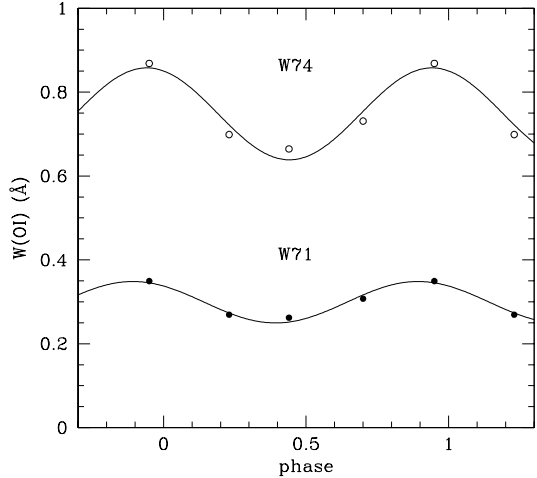}
\caption{W71 and W74 curves of the Cepheid SS Sct. The phases were calculated using a
 period of P = 3.671280d and epoch of maximum light
 hjd 2444398.419. The amplitudes of 
 W71 and M74 variations are 0.098 and 0.220 angstroms respectively.}  
\end{figure}

\begin{figure}  
\includegraphics[width=7.cm]{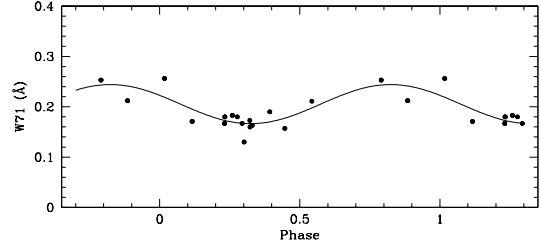}  
\caption{W71 curve of the Cepheid $\zeta$ Gem. The phases were calculated using a
 period of P= 10.15078d  and epoch of maximum light
 hjd 2444232.443. The amplitude of the  W74 variation is 0.078 angstroms.}
\end{figure}

Since the luminosity of a Cepheid changes as the star pulsates, the values of W71 and W74
also change along the cycle, thus W71($\phi$) and W74($\phi$) at a given phase $\phi$ 
must be corrected to be brought to the mean values of W71 and W74.
 Knowing the lines contributing to the OI triplet are
    luminositiy sensitive, it is reasonable to assume that the
amplitudes of the W71- and W74-curves are proportional to the V-light curve 
amplitude.       
 Thus, the size of the equivalent width correction to be applied to bring it
 to the mean value 
depends on the phase of the observation and the V-light amplitude. 
To properly estimate the scale between V-light and W74 and W71 variations amplitudes, one
should meassure W71 and W74 at several phases  for a group of Cepheids.
 However, due to observing time limitations we
were able to meassure W71 and W74 only at a few phases in SS Sct. The variations
are shown in Fig. 3, where we have used the ephemeris $\phi_i = (t_i - 2444398.419)/3.671280$

\begin{table*}
\caption{OI 7774 data for Classical Cepheids}                                                                 
\begin{center}
\begin{tabular}{lccccccc}
\noalign{\smallskip}                                                              
\hline       
\noalign{\smallskip}                                                              
\noalign{\smallskip}
\multicolumn{1}{l}{Name}&
\multicolumn{1}{c}{W71($\phi$)}& 
\multicolumn{1}{c}{W74($\phi$)}& 
\multicolumn{1}{c}{P}& 
\multicolumn{1}{c}{$M_V$}&
\multicolumn{1}{c}{$\phi$}&
\multicolumn{1}{c}{W71$_o$}&
\multicolumn{1}{c}{W74$_o$}\\
\noalign{\smallskip}
\multicolumn{1}{l}{}&
\multicolumn{1}{c}{($\AA$)}& 
\multicolumn{1}{c}{($\AA$)}& 
\multicolumn{1}{c}{(days)}& 
\multicolumn{1}{c}{(mag.)}&
\multicolumn{1}{l}{}&
\multicolumn{1}{c}{($\AA$)}&
\multicolumn{1}{c}{($\AA$)}\\
\noalign{\smallskip}
\hline  
\noalign{\smallskip}

DT Cyg & 0.268 & 0.679 & 2.499035&-2.548&  0.45&0.278&0.702\\
V532 Cyg &0.320 & 0.826 &3.283612& -2.881 & 0.92&0.289&0.755\\
SS Sct  &  0.262& 0.665 & 3.671280& -3.017 & 0.44&0.280&0.718\\
RT Aur & 0.187 & 0.447 & 3.728220&-3.036 & 0.61&0.238&0.563\\
SU Cyg & 0.205 & 0.485 &3.845733& -3.074 & 0.40&0.240&0.564\\
CM Sct &  0.218 & 0.566 &3.916977&-3.096 & 0.60&0.257&0.653\\
BQ Ser & 0.251 &  0.632&4.316700& -3.215 & 0.74&0.247&0.621\\
T Vul  &   0.165 &  0.477 & 4.435532& -3.248&0.65&0.247&0.722\\
VZ Cyg & 0.259 & 0.687&4.864504& -3.361&  0.19&0.226&0.611\\
V Lac  &  0.289 &0.716&4.983149& -3.390 & 0.17&0.232 &0.587\\
AP Sgr & 0.198&  0.542 & 5.057936&-3.408 & 0.75&0.295&0.760\\
V350 Sgr& 0.193 & 0.583&5.154557& -3.431 & 0.51&0.234&0.741\\
V386 Cyg & 0.338&  0.946&5.257655 &-3.455&  0.90& 0.314&0.894\\
$\delta$ Cep& 0.242 & 0.577 &5.366316& -3.480&0.32&0.252&0.633\\
X Lac &   0.209 & 0.561&5.444990&-3.498&  0.26&0.208 &0.557\\
Y Sgr &  0.239& 0.589&5.773400& -3.570 & 0.34&0.238&0.588\\
FM Aql & 0.314 & 0.838 &6.114240& -3.640 & 0.08&0.232&0.656\\
X Vul & 0.219& 0.655 &6.319562& -3.680 & 0.70&0.344&0.932\\
RR Lac &0.219 & 0.599& 6.416190&-3.698 & 0.75&0.295&0.771\\
AW Per & 0.190 & 0.470&6.463589& -3.707&  0.70&0.267&0.645\\
U Aql &  0.362 & 0.921 &7.024100&-3.809 & 0.91&0.322&0.830\\
$\eta$ Aql & 0.228& 0.519 &7.176779&-3.835 &0.67&0.303&0.687\\
V600 Aql& 0.298 &  0.471 & 7.238748&-3.846&0.71&0.254&0.652\\
V459 Cyg& 0.160 & 0.444&7.251250&-3.848&  0.71&0.231&0.602\\
W Sgr &0.210 &  0.532 &7.595080& -3.904&0.55&0.294&0.760\\
U Vul & 0.343 & 0.930 & 7.990736&-3.966&0.92&0.296&0.829\\
S Sge & 0.270 & 0.680 & 8.382044&-4.025 & 0.31&0.254&0.644\\
YZ Sgr & 0.291 & 0.738 & 9.553606&-4.184&0.63&0.305&0.769\\
Y Sct & 0.306&0.822 &10.341650&-4.281& 0.05&0.224&0.639\\
TT Aql& 0.316 & 0.842&13.755290& -4.629&0.92&0.287&0.777\\
CD Cyg & 0.344& 0.764&17.073967&-4.893 & 0.85&0.420&0.935\\
            \noalign{\smallskip}
            \hline  
            \noalign{\smallskip}  
\end{tabular}
\end{center}

\end{table*}  

\begin{figure}[t]  
\includegraphics[width=7.cm]{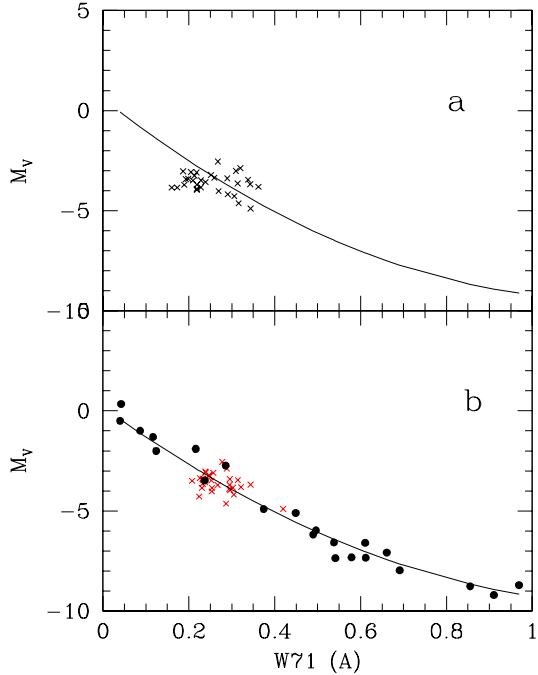}
\caption{Distribution of Cepheids in the W71 - $M_V$ plane. The solid line
is the calibration in Figure 1b obtained from the calibrators in Table 1. a) Distribution of 
Cepheids before W71 correction.
b) The small crosses are the Cepheids positions after their W71 values were corrected from pulsational
phase, see text for details.}  
\end{figure}

\begin{figure}[t]  
\includegraphics[width=7.cm]{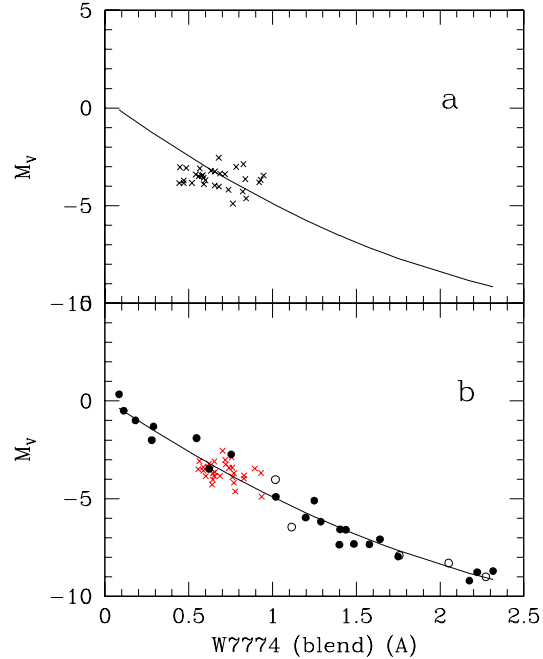}
\caption{Same as Fig. 4. for W74. Open circles as in Fig. 1.}  
\end{figure}

For a sample of Cepheids we can represent the light curve using the
 Fourier coefficients calculated by Arellano Ferro et al. (1998).
  The expression  used is of the form,
\begin{equation}
V(\phi) = V_o + \sum_{k=1}^{n}{A_k cos(2\pi k\phi +\Phi_k),}
\end{equation}

\noindent where $A_k$ and $\Phi_k$ are the amplitude and the displacement of each harmonic $k$.
Thus we can calculate $\Delta V(\phi) = V(\phi) - V_o$. We now assume that
$F_{74} = A_V/A_{W74}$, i.e. the ratio of the V light amplitude to the W74 variation amplitude
 $A_{W74}$, is constant over the complete cycle, or
\begin{equation}
F_{74} = A_V/A_{W74} =\Delta V(\phi)/\Delta W74(\phi),
\end{equation}

\noindent
where $\Delta \rm{W74}(\phi) = \rm{W74}(\phi)-\rm{W74}_o$.
Then the estimated mean value of W74 would be given by
\begin{equation}
W74_o=W74(\phi) - \Delta V(\phi)/F_{74},
\end{equation}

\noindent
which can be used to estimate the mean absolute magnitude for a cepheid
from equation 2. Similar arguments hold for W71 and eq. 1.

The estimated $F$-values for the reference star SS Sct are $F_{71}=4.388 
$ and $F_{74}=1.955$. The uncertainties of these values are proportional 
to the scatter of both V-light and W(OI) curves. In Figure 4 we have presented
the W71 curve for the cepheid $\zeta$ Gem for which W74 curve is very noisy.
Since this star has large variation in temperature and during the
cooler phase the contributions from the Fe I and CN features mentioned
 in section 3 are large,
 hence the errors in W74 also become large. For this star we
 calculate $F_{71}=6.01$ and it would therefore produce similar corrections
to those using SS Sct. For the other Cepheids corrections, we have
used SS Sct as a reference for both W71($\phi$) and W74($\phi$).
 
It should be noted that the above approach may have limitations since
 the amplitude scale factor calculated in this manner may not be applicable
 to Cepheids with highly asymmetrical light curves. However, this should
be considered as a maiden effort with considerable room for improvement.
 We intend carrying out a similar calculation for a large sample of Cepheids.

This approach to random phase correction in Cepheids is
nevertheless presented here as a preliminary result and as a 
promising method to estimate  the $M_V-W(OI~7774)$ from random-phase
 observations of Cepheids.

\begin{figure*}[t]  
\includegraphics[width=16cm,height=8cm]{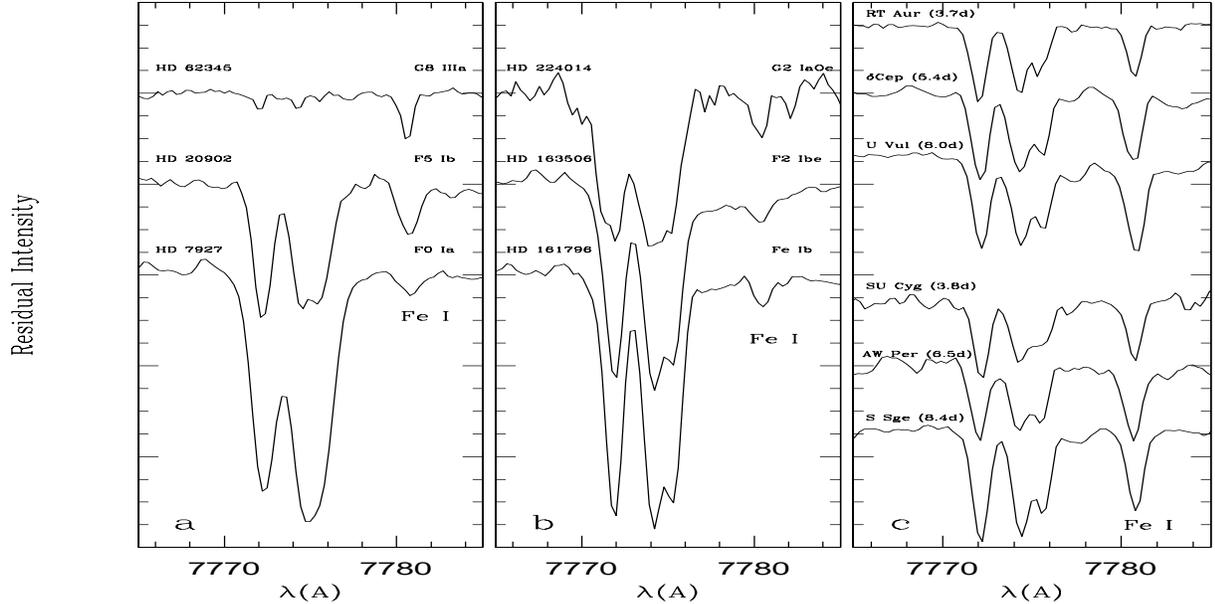}
\caption{a) Three examples of calibrator stars from very low to very large luminosity. b) Three examples of evolved stars. These stars known for having H$\alpha$
in emission do not show emission in the OI7774 feature. Other stars in Table 3 were inspected and no emission was found, thus their values W71 and W74 reliable for
the estimation of $M_v$. c) The three upper spectra are from single Cepheids while the three examples in the bottom are of Cepheids
known for having hot companions (see Table 2 of Evans (1995) for details). Numbers between parenthesis are their periods in days. 
The presence of the companion does not seem to affect the nature of the OI7774 in Cepheids.}  
\end{figure*}

The above method has been applied to both W71 and W74 data of 31
 Cepheids, listed
in Table 2 along with their randon phase W71($\phi$), W74($\phi$), 
$M_V$ and the mean values W71$_o$ and W74$_o$. The $M_V$ values were obtained
from the Feast \& Catchpole (1997) P-L relationship
$M_V = -2.81~ log~ P - 1.43$. 
This calibration is brighter than other solid calibrations (e.g. Sandage
\& Tammann 1968; Feast \& Walker 1987; Madore \& Freedman 1991), but only at 
the level of  $\sim$ 0.1 mag (Sandage\& Tammann 1998; Madore \& Freedman 1998), thus,
adopting a calibration of the P-L relationship with a slightly smaller zero point, would have a very minor effect on our calculations and conclusions, especially considering that the uncertainty of
our OI7774-$M_V$ calibration is of the order of 0.4 mag.
The periods and epochs are usually known with
large precision, the values in Table 2 were adopted from
the sources listed in the paper by Arellano Ferro et al. (1998)
 in their Table 2. In the present  Table 2, we include the phase
at which the OI observation was obtained and, on the basis of which, the
absolute magnitude is corrected.

Fig. 5a shows the distribution of uncorrected W71 meassurements in the
W71-$M_V$ plane. The solid curve is the calibration in equation 1. While
the Cepheids fall
along the path defined by the non-cepheid calibrators, their dispersion is, as
expected, unacceptably large as the phase effect is yet to be corrected.
In Fig. 5b the corrected W71-values for the sample of Cepheids are plotted
along with the non-cepheid calibrators. The process has been repeated for W74 and
is shown in Fig. 6. The present corrections of W71 and W74 in
fact brought   the Cepheids closer to the mean trend, and their dispersion is now
comparable to that of cluster, associations and field calibrators. Once the
Cepheids are included the calibrations take the forms:

\begin{equation}
\begin{array}{@{}r@{}r@{~}r@{~}r@{~}r@{~}r@{}}
M_V = &\phantom{\pm}0.260& - 15.889&\rm{W71}& + 6.393&\rm{W71}^2, \\
 &\pm0.281&\phantom{0}\pm1.387& &\pm 1.390& 
\end{array}
\end{equation}

\begin{equation}
\begin{array}{@{}r@{}r@{~}r@{~}r@{~}r@{~}r@{}}
M_V = &\phantom{\pm}0.131& - 5.831&\rm{W74}& + 0.789&\rm{W74}^2. \\
 &\pm0.287&\phantom{0}\pm0.563& &\pm 0.231& 
\end{array}
\end{equation}

The standard deviations in $M_V$ are 0.42 and 0.43 mag. thus comparable
to the calibrations in eqs. 1 and 2.

Many Cepheids have hot companions, generally late B-type main sequence stars,
And while contamination of the cepheid OI7774
Feature is unlikely since hot main sequence stars have very weak 
OI7774 relative to supergiant stars (Faraggiana et al. 1988),
We have compared in figure 7-c the  OI7774 profiles of three Cepheids with B-A
type main sequence companions (SU Cyg, AW Per and S Sge) with
those of presumably single Cepheids of similar periods (RT Aur,
$\delta$ Cep, and U Vul). We do not see any peculiarity introduced by the companions,
hence it is concluded that the
OI7774 feature in Cepheids is not affected by their companions.

\begin{table}[t]
\footnotesize{
\caption{OI 7774 data and luminosities of selected evolved stars and
established post-AGB stars.}                                                                 
\begin{center}
\begin{tabular}{rlcccc}

\noalign{\smallskip}                                                              
\hline       
\noalign{\smallskip}
\noalign{\smallskip}
\multicolumn{1}{r}{HD}&
\multicolumn{1}{l}{Sp.T.}& 
\multicolumn{1}{c}{W71}& 
\multicolumn{1}{c}{W74}& 
\multicolumn{1}{c}{$<M_V>$}&
\multicolumn{1}{c}{$log L/L_{\odot}$}\\
\noalign{\smallskip}
\hline  
\noalign{\smallskip}  
725    &F5Ib-II& 0.450 & 1.220  &--5.723 & 3.905 \\
1457   &F0Iab& 0.461 & 1.198  &--5.737 & 3.927\\
4266   &F2Iab& & 1.028     &--5.023 & 3.633\\
9167   &F1II& 0.544 &1.397    &--6.530 & 4.240 \\
9233   &A4Iab&  0.489 &1.328    &--6.136 & 4.142 \\
12533  &K3IIb&  0.070 & 0.146  & --0.510&2.008  \\
12545  &G5& 0.148&   0.345 &--1.691 & 2.316 \\
15257  &F0III&  &0.655   &--3.267 &2.939   \\
15788  &G8III& 0.048 &0.129  &--0.281 &1.772 \\
27381  &F2& 0.505&  1.343   &--6.244 &4.121 \\
54605  &F8Iab:&0.691 & 1.750  &--7.726 &4.710 \\
55612  &F0III/IV& 0.068& 0.211  &--0.686 &1.906 \\
55661  &A7:V& 0.198& 0.558  &--2.632 &2.705 \\
57321  &F2II& 0.661&   1.027  & --6.273&4.133 \\
61227  &F0Ib&  0.148 & 0.383  &--1.796 &2.343 \\
62058  &F8/G0Ia& 0.456& 1.144  &--5.599 &3.864 \\
137569 &B5III& 0.290 & 0.804    &--3.870 &3.704 \\ 
191635 &F0& 0.234 &  0.610  &--3.018 &2.839 \\
194093 &F8Iab:& 0.436 &   1.149   &--5.504 &3.821 \\
202240 &F0III& 0.322 &0.987  &--4.494 &3.430 \\
209747 &K4III&0.046& 0.087&--0.137&1.899\\
216756 &F5II& 0.172 &0.477  &--2.230 &2.508 \\
224014 &G2Ia0e& &1.468  &--6.777 &4.340 \\
            \noalign{\smallskip}  
            \hline  
            \noalign{\smallskip}
112374 &F3Ia& &1.026&--5.014&3.630$^1$ \\
&& &&$(-4.44)$& \\
161796 & F3Ib & 0.747 &  2.008 &--8.278 &4.935 \\
&& &&$(-8.5)^2$& \\
163506&F2Ibe& 0.634 & 1.663  &--7.375 & 4.574\\
&& &&$(-8.1)^2$& \\
172324& B9Ib&0.440 & 1.153&--5.533 &4.021$^3$\\
&& &&$(-5.98)$& \\
172481 &F2Ia0& 0.458 &   1.121  &--5.562 & 3.849$^3$\\
&& &&$(-6.44)$& \\
            \noalign{\smallskip}
            \hline  
            \noalign{\smallskip}  
\end{tabular}
\end{center}
Notes. 1. Luck et al. (1983), 2. Arellano Ferro \& Parrao (1990), 3. Arellano Ferro et al. (2001).
}
\end{table}

\section{Esimated Luminosities for AGB candidate  stars}
\label{sec:Luminosities}

 We felt it would be important to estimate the $M_V$ of possibly
 evolved objects from their W71 and W74
 using the calibration derived in the present work.
 In a different program we had chosen a sample of A-G stars with
 high galactic latitude and detected
 IR flux in search of post-AGB stars. Not all of them turned out
 to be objects showing very significant chemical peculiarities caused by
 evolutionary processes. But we were interested in determining their
 locations in the H-R diagram, hence used their W71, W74 to estimte
 $M_V$ for them. Their temperatures are those estimated using 
 fine spectral analysis as described in Arellano Ferro, et al. 2001.
  In absence of such data we relied upon $uvby\beta$
 or 13-colour photometry calibrations or their spectral types.
The stars under consideration are listed in Table 3 along with
 their W71 and W74 values. Also reported in Table 3 is $<M_V>$,
 the mean of the absolute magnitudes obtained from eqs. 1 and 2 which
agree within 0.2 mag. The corresponding $log~L/L_\odot$ is also given
in Table 3. At the bottom of the table, we have presented the data for five
 well-established post-AGB stars and their  OI derived luminosities. Previous $M_V$-values are given
between parethesis and their sources are listed in Table 3.

The OI7774 profiles of evolved stars in table 3 were inspectes to make sure that
Equivalent widhts were not affected by emission often present in evolved or unusual stars. To demonstrate that it was not the case, in Fig. 7-B three spectra, of stars HD 224014 ($\rho$ Cas, G2Ia0e),
HD 163506 (89 Her, F2Ibe) and HD 161796 (F3Ib) are shown.
These stars were selected for illustration since they are well known to have H$\alpha$
In emission. Nevertheless, the OI7774 feature appears to be in absorption. There could be
weak underlying emission, but its effect, if any, would be negligible.

The positions of the stars on the H-R diagram are shown in Fig. 8.
The evolutionary tracks for several masses of Schaller et al. (1992) for Z=0.02 and Y=0.30
are presented for reference. These models do not show blue loops for stars with 4 $M_{\odot}$
and below, hence do not cross the instability strip. On the other hand, lower metallicity
models (Z=0.001)
do show blue loops for masses down to about 2-3 $M_{\odot}$ (Schaller et al. 1992) and
cross the lower part of the cepheid
instability strip. While such selection of metallicity would be adequate for old low mass stars,
it would be inappropriate for Pop I Cepheids. To produce longer blue loops at this low mass
range, Alongi et al. (1991) have introduced an extra overshoot parameter that extends towards
the interior
the outer convective envelope. Post red giant branch stars located on the H-R diagram
might serve as land-marks for theoretical work.

The stars from the Table 3 with $log~L/L\odot$ in the range of 3.5 to 5.0 have certainly passed the red giant phase and populate the blues loop for masses betwee 7 and 10 $M_{\odot}$. Blue loops for higher masses are not populated in part due to the specific sample considered but also due to the fact that evolution in this region of the diagram is fast (Bl\"ocker 1995). However it is not possible to say if a given star would evolve to the left or to the right. Stars like HD 137569 and HD 172324 might be useful observational input that can be used to examine the extent of blue loops in this mass range.

The five established post-AGB stars, plotted as dots in Fig. 8, have been
plotted according to their OI luminosities and their spectroscopically determined
temperatures. Although the OI7774 feature is sensitive to the luminosity,
it is also partially sensitive to the oxygen abundance. The calibrations
have been established using nearly solar abundance calibrators.
Therefore its application to highly evolved stars with peculiar
oxygen abundances is  a little uncertain and hence might give luminosities with large
 error bars.
The oxygen [O/H] abundances for these five post-AGB are:
--0.33 (HD112374; Luck et al. 1983), +0.08 and --0.27 (HD 161796 and HD163506;
Luck et al. 1990),  +0.41 and --0.58 (HD 172324 and 172481; Arellano Ferro et al.
2001), therefore their positions on the H-R diagram may
 have larger uncertainties.
 Mildly evolved stars like those given in the upper part of Table 3
 have essentially solar [O/H], hence the calibration certainly gives
 good $M_V$ estimates for them. For C-rich strongly evolved objects
 the relation might give low values of the luminosity.

Also as a reference, the position of the instability strip has been indicated in Fig 8. The upper strip is the classical cepheid strip from Sandage  \& Tammann  (1969) and the lower strip comes from Marconi \& Palla (1998). Given the 0.4 mag. uncertainty in $M_v$ produced by eqs. 1 and 2, we cannot assure that borderline
Cases are in or out the instability strip. It is worth however pointing at the variables sitting well inside the strip, 
HD 112374 whose variability was discivered by Arellano Ferro (1981),
HD62058 (R PUP) and HD 194093 (37 Cyg). Variable stars in the upper part of the diagram, like HD 54605 (25 CMa), HD 163506 (89 Her), HD 161796 (V814 Her) and HD224014 ($\rho$ Cas) lie in a region where the instability strip is ill-defined. 

\section{Conclusions}
\label{sec:Conclusios}

The newly estimated values of the absolute magnitude $M_V$
 for a group of selected
 A-G supergiant calibrators enabled us to revise the $M_V-W(OI~7774)$ relationship.
 The results show
an improvement in $M_V$ predictions accuracies of at least a factor of three relative
to the previous calibration from high resolution data in paper I, over a large
absolute magnitude range --9.5 to +0.35 mag. The calibrations presented in equations
 1 and 2 are our final calibrations and 
they predict M$_V$ values with an accuracy of $\pm$0.38 mag.

\begin{figure}[t]
\includegraphics[width=7.cm]{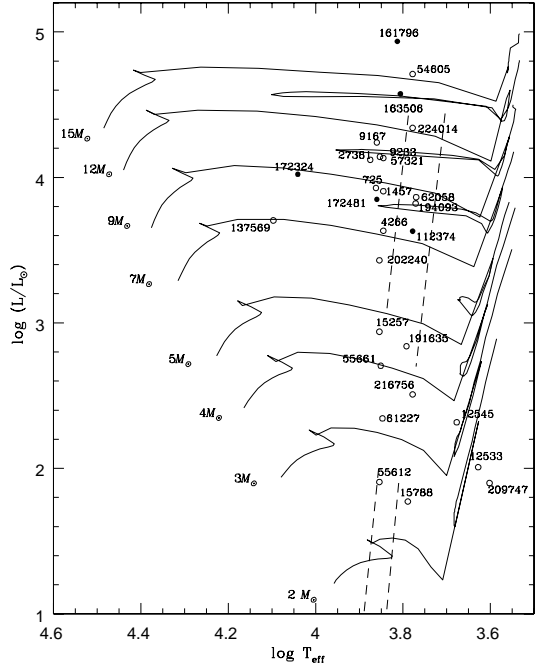}
\caption{H-R diagram with the positions of selected evolved stars (open circles) and five
established post-AGB stars (dots). The evolutionary
tracks are from Schaller et al. (1992) for Z=0.2  and Y =0.30.}
\end{figure}

It is shown that the OI7774 feature in Cepheids follow the basic calibration. A method
to correct $W(OI~7774)$ obtained at random phases is described and succesfully
applied to a sample of 31 Cepheids.
At our resolution, the bluemost component of the triplet at
$\lambda 7771.954$ is resolved from the two redder components
 and being unaffected by FeI, CN lines may be more useful for 
G-type stars and later.
We have calculated the calibrations for
both the blue component (W71) and for the blend of the triplet (W74) including 
the Cepheids along with the primary calibrators. The calibrations for composite data, given by eqs. 8 and 9, 
predict M$_V$ values within $\pm$0.42 and $\pm$0.43 mag. and therefore are 
 comparable
 to the calibrations based solely on A-G non-variable supergiants. Hence the present
 calibration with phase-corrected W71 or W74 shows that the OI7774 feature in Cepheids
 is as sensitive to luminosity as in non-variable supergiants

The new calibrations have been applied to a group of intermediate
 temperature, high galactic
latitude stars  with detected IR fluxes that                    
are considered good candidates to post-AGB stars.
The luminosities determined by the present work, not only help in
ascertaining the evolutionary status of the sample of stars
but can also be used by theorist doing evolutionary calculations in the
post red giant evolution,
in order to establish the loci of the blue loops for stars of five solar masses and below.

\acknowledgments

AAF acknowledges support from DGAPA-UNAM grant through project IN113599. 
SG is thankful to the Department of Science and Technology, India,
for the travel support under Indo-Mexican Collaborative project
DST/INT/MEXICO/RP001/2001. We are grateful to an anonymous referee
for helpful suggestions. This work has made a large use of the SIMBAD and ADS
services, for which we are thankful.

\end{document}